\documentclass[sigconf]{acmart}

\setcitestyle{sort&compress}
\AtBeginDocument{}

\usepackage{graphicx}
\usepackage{subcaption}
\usepackage{tikz}
\usepackage{arydshln}
\usepackage{svg}
\usepackage{xcolor}
\usepackage{makecell}
\usepackage{float}
\usepackage{booktabs}

\copyrightyear{2026}
\acmYear{2026}
\setcopyright{cc}
\setcctype{by}
\acmConference[CODASPY '26]{Proceedings of the Sixteenth ACM Conference on Data and Application Security and Privacy}{June 23--25, 2026}{Frankfurt am Main, Germany}
\acmBooktitle{Proceedings of the Sixteenth ACM Conference on Data and Application Security and Privacy (CODASPY '26), June 23--25, 2026, Frankfurt am Main, Germany}
\acmDOI{10.1145/3800506.3803508}
\acmISBN{979-8-4007-2562-3/2026/06}

\title[SecRL-Prune]{SecRL-Prune: Structured Reinforcement Learning–Based Pruning of CodeLLMs for Preserving Adversarial Code Mutation}
\author{Parsa Memarzadehsaghezi}
\affiliation{%
  \institution{Ontario Tech University}
  \city{Oshawa}
  \state{Ontario}
  \country{Canada}
}
\email{parsa.memarzadeh@ontariotechu.ca}

\author{Pooria Madani}
\affiliation{%
  \institution{Ontario Tech University}
  \city{Oshawa}
  \state{Ontario}
  \country{Canada}
}
\email{pooria.madani@ontariotechu.ca}

\author{Khalil El-Khatib}
\affiliation{%
  \institution{Ontario Tech University}
  \city{Oshawa}
  \state{Ontario}
  \country{Canada}
}
\email{khalil.el-khatib@ontariotechu.ca}

\renewcommand{\shortauthors}{Parsa Memarzadehsaghezi, Pooria Madani, and Khalil El-Khatib}
\keywords{Code Large Language Models, Structured Pruning, Reinforcement learning, Code mutation, Metamorphic malware, Malware detection}

\begin{document}

\begin{abstract}
Large code language models (CodeLLMs) can generate and rewrite programs, enabling functionality-preserving code mutation that may be used to create diverse malware variants and evade signature-based detection. A key security question is whether this mutation capability survives model compression, which would make deployment feasible under limited hardware budgets. We propose \textsc{SecRL-Prune}, a structured pruning framework for CodeLLMs that operates on feed-forward (MLP/FFN) channels. Starting from a pretrained teacher, it learns a layer-wise pruning policy with reinforcement learning using a teacher--student KL-divergence reward. To improve efficiency, we cache the teacher's top-$P$ predictions once and compare the pruned student against this compact target, avoiding simultaneous teacher--student residency in GPU memory. We evaluate \textsc{SecRL-Prune} on HumanEval using $\mathrm{pass}@k$ for execution correctness and $\mathrm{var}@k$ for code diversity across three 7B CodeLLMs at 10--30\% compression. \textsc{SecRL-Prune} consistently preserves higher $\mathrm{pass}@k$ and $\mathrm{var}@k$ than recent structured pruning baselines under aggressive pruning. In a case study on real malware samples, semantics-preserving mutations from 20\%-pruned models substantially reduced detections. These results show that code mutation capability can survive significant structured pruning, highlighting the security relevance of compressed CodeLLMs.
\end{abstract}

\ccsdesc[500]{Security and privacy~Malware and its mitigation}
\ccsdesc[500]{Computing methodologies~Reinforcement learning}
\maketitle
\section{Introduction}

The emergence of large language models capable of code generation---commonly termed \emph{CodeLLMs}---has fundamentally transformed numerous domains within computer science. These models offer substantial benefits for accelerating software development and streamlining maintenance workflows; however, they simultaneously present significant security concerns. Threat actors can exploit the same generative capabilities to author malware, craft software exploits, and automate vulnerability discovery. Recent threat intelligence reports document adversaries employing LLMs to obfuscate malicious code, generate context-aware payloads, and orchestrate sophisticated intrusion campaigns, confirming that AI-augmented offensive tooling has transitioned from theoretical concern to operational reality~\cite{lemos2025llmmalware,anthropic2025aiespionage}. Of particular concern is the potential for adversaries to leverage CodeLLMs' rapid generation capabilities to systematically mutate existing malware, thereby evading signature-based detection mechanisms while continuously reintroducing new variants into the threat landscape.

This emerging application of CodeLLMs reflects a well-established lineage of code mutation techniques in sophisticated malware~\cite{wang2019metahunt,wong2006hunting,madou2006software,konstantinou2008metamorphic}. Because many detection systems continue to rely on static signatures---whether binary patterns, syntactic structures, or system call sequences---malware authors are strongly incentivized to transform their code's representation while preserving its underlying functionality~\cite{bensaoud2024survey,wang2019metahunt,konstantinou2008metamorphic}. Historically, such transformations have been achieved through automated, rule-based mutation engines compact enough to be embedded directly within the malicious payload. This architecture enables real-time polymorphism as the malware propagates through operational environments. Metamorphic malware exemplifies this approach: by rewriting its own code with each replication cycle, it has proven remarkably effective at circumventing static analysis methods~\cite{wong2006hunting,madani2024metamorphic}.

Contemporary CodeLLMs already facilitate offline variant generation, wherein attackers leverage separate infrastructure to produce novel malware strains prior to deployment. However, a more consequential threat lies in the prospect of miniaturization---compressing these models sufficiently to embed directly within malware payloads or execute rapidly on consumer-grade hardware~\cite{setak2024finetuning,madani2024metamorphic}. The central technical question is whether modern CodeLLMs can be reduced to a minimal footprint through pruning, distillation, or quantization while retaining their capacity for semantics-preserving code transformation/mutation~\cite{dettmers2023qlora,frantar2023sparsegpt,wei2024effectively}. Crucially, many auxiliary capabilities---such as documentation generation, test synthesis, and general-purpose prompt-based programming---become superfluous when the model's sole objective is code mutation. This observation motivates the hypothesis that specialized, highly compressed models optimized exclusively for code mutation are not only feasible but represent a significant and understudied attack vector~\cite{setak2024finetuning,madani2024metamorphic}. Our work directly addresses this gap by systematically investigating the compressibility limits of CodeLLMs with respect to their code mutation fidelity.

In this work, we take a first step toward rigorously evaluating this hypothesis. Our key contributions are as follows:
\begin{enumerate}
  \item We propose \textsc{SecRL-Prune}, an RL-based structured pruning framework for CodeLLMs that uses a KL-guided pruning objective to preserve teacher behavior under compression.
  \item We introduce a caching mechanism that accelerates RL training and reduces the resource footprint of pruning: by precomputing teacher outputs on a calibration set, \textsc{SecRL-Prune} can train its policy with only the student model in memory, yielding lower peak GPU usage than prior structured pruning techniques.
  \item We conduct a systematic empirical study of how structured compression affects CodeLLMs' ability to perform functionality-preserving, syntax-diverse code mutation.
  \item  We conduct a case study using real malware samples to assess how semantics-preserving code mutations (using pruned CodeLLMs) affect malware detection across more than $62$ industry-standard malware detection engines.
\end{enumerate}

The rest of this paper is structured as follows: In Section 2, we provide a survey of the existing body of work on CodeLLMs and code mutation literature. In Section 3, we provide the necessary background on the structure of LLMs that we leverage in our pruning technique and discuss the fundamentals of how our reward function is designed. In Section 4, we present SecRL-Prune, our novel approach to pruning CodeLLMs. In Section 5, we provide a detailed description of our experimental setup, including the datasets used. In Section 6, we discuss the results, and in Section 7, we provide open questions and a path forward for this research.

\section{Related Work}
Malicious use of \emph{CodeLLMs}, particularly for malware code mutation, has been investigated in recent work on LLM-driven code mutation and metamorphic malware evolution. These studies~\cite{setak2024finetuning,madani2024metamorphic} provide both survey-style overviews and empirical analyses of the code-mutation capabilities of open- and closed-source CodeLLMs. In contrast, our work focuses on recent advances in LLM compression, with an emphasis on preserving functionality relevant to adversarial code generation/mutation~\cite{sanh2019distilbert,frantar2023sparsegpt,wei2024effectively}. 

Model compression aims to reduce model size and improve transformer efficiency while preserving task performance (e.g., code mutation capabilities), using techniques such as knowledge distillation~\cite{sanh2019distilbert,wang2020minilm,gu2024minillm}, quantization~\cite{dettmers2023qlora}, and pruning~\cite{frantar2023sparsegpt,ma2023llmpruner}.

In knowledge distillation, a smaller student LLM is trained to mimic the predictions or internal signals of a larger teacher, typically using a Kullback-Leibler(KL) divergence--based loss between teacher and student distributions~\cite{kullback1951information}. DistilBERT compresses BERT via such an approach, achieving roughly half the parameter count while retaining most of BERT's language understanding, by combining supervised cross-entropy with a KL-divergence loss on softened logits~\cite{sanh2019distilbert}. MiniLM focuses on distilling self-attention knowledge, training the student to match the teacher's attention distributions and value relations via KL divergence~\cite{wang2020minilm}, and MiniLLM transfers the behaviour of large generative LLMs into smaller students using an objective tailored for generation~\cite{gu2024minillm}.

Quantization techniques, by contrast, reduce the numerical precision used to represent LLM weights to save memory and accelerate inference. For example, QLoRA combines 4-bit quantization with low-rank adapters to enable memory-efficient fine-tuning of large models on standard hardware~\cite{dettmers2023qlora}.

Pruning directly removes parameters from an already trained network. In unstructured pruning, individual weights are zeroed out rather than physically removed from the model. This produces fine-grained sparsity patterns that can be highly parameter-efficient in theory, and techniques like SparseGPT use blockwise approximations of the Hessian to decide which weights can be zeroed in LLMs~\cite{frantar2023sparsegpt}. However, because the original tensor shapes are preserved and zeros are scattered irregularly, practical speedups often depend on specialized sparse kernels or hardware support. Without such support, the compressed model may still execute as a dense network.

In structured pruning, the basic units being removed are larger groups of weights—such as channels, neurons, attention heads, or even whole layers—so that tensor dimensions themselves shrink, yielding architectures that are inherently easier to deploy. Early work in this direction, such as LLM-Pruner, scores "coupled" structures (e.g., head–MLP groups) using gradient and curvature information, removes low-importance groups, and applies lightweight fine-tuning (e.g., LoRA-style adapters) to recover accuracy~\cite{ma2023llmpruner}. Subsequent methods have pushed structured pruning further in two directions. One line of work explores runtime adaptivity, dynamically adjusting which structures are pruned during inference to trade off cost and quality on the fly~\cite{liu2025rap}. Another formulates pruning as policy learning, training a reusable pruning policy that generalizes across compression ratios without requiring additional calibration data~\cite{sengupta2025you}.

Overall, these techniques show that both unstructured and structured pruning can substantially reduce the effective parameter count of large language models, and that recent post-training methods can achieve high compression ratios with modest fine-tuning. However, most of this work targets natural-language LLMs and is evaluated on generic language understanding or generation benchmarks. Their behaviour on code-generation models and, in particular, their impact on execution-based metrics and mutation/diversity properties of CodeLLMs remains comparatively underexplored. Recall that such compression of CodeLLMs can be an important enabling factor in the development of next-generation metamorphic malware, which is the core thesis of this manuscript.

\section{Background}

Before introducing our novel approach for CodeLLMs pruning, we first review the transformer architecture underlying modern CodeLLMs and introduce the notation and Kullback–Leibler (KL) divergence used in our work. 

\subsection{Transformer Structure in CodeLLMs}
\label{sec:mlps}
\begin{figure}
    \centering
    \includegraphics[width=1.2\linewidth]{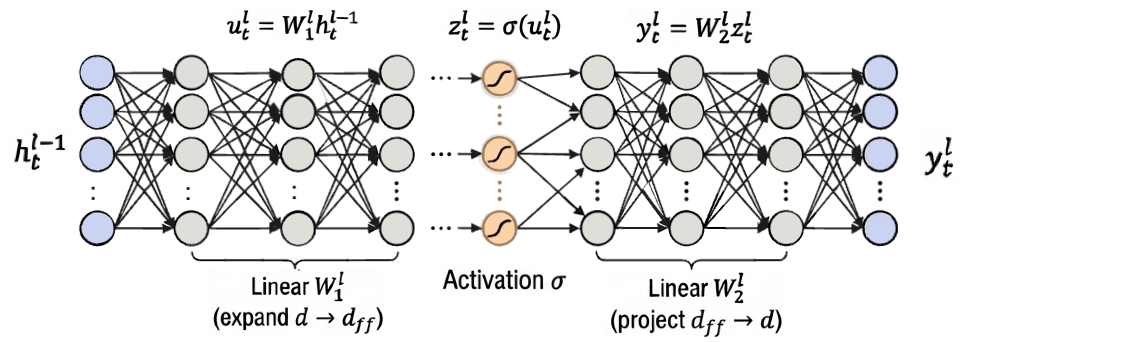}
    \caption{Feed-forward (MLP/FFN) sub-layer in a transformer block. Given the input hidden state $h_t^{l-1}\in\mathbb{R}^d$, the first linear layer $W_1^l$ expands to an intermediate width $d_{\mathrm{ff}}^l$, a nonlinearity $\sigma(\cdot)$ produces activations $z_t^l$, and the second linear layer $W_2^l$ projects back to $\mathbb{R}^d$ to yield $y_t^l$. Each coordinate of $z_t^l$ corresponds to an MLP channel, which is the structured unit pruned by \textsc{SecRL-Prune}.}

    \label{fig:mlp}
\end{figure}

In this work, our pruning actions operate directly on the intermediate \emph{channels} of the feed-forward network (FFN) inside each transformer block. We therefore focus our discussion on this component. Figure~\ref{fig:mlp} shows the standard two-layer FFN used in decoder-only transformers.

Consider transformer block (layer) $l$. For a token position $t$, let $h_t^{\,l-1}\in\mathbb{R}^d$ be the input hidden state entering the MLP (the $d$-dimensional residual stream). The MLP applies two linear layers with a pointwise nonlinearity in between:
\begin{equation}
u_t^l = W_1^l\, h_t^{\,l-1}, \qquad
z_t^l = \sigma(u_t^l), \qquad
y_t^l = W_2^l\, z_t^l,
\end{equation}
where $W_1^l \in \mathbb{R}^{d_{\mathrm{ff}}^l \times d}$ expands the representation from hidden size $d$ to an intermediate width $d_{\mathrm{ff}}^l$, $\sigma(\cdot)$ is a pointwise activation function (e.g., GELU), and $W_2^l \in \mathbb{R}^{d \times d_{\mathrm{ff}}^l}$ projects back to $\mathbb{R}^d$. The output $y_t^l$ is then added back to the residual stream via the standard residual connection.

The intermediate activation $z_t^l \in \mathbb{R}^{d_{\mathrm{ff}}^l}$ has $d_{\mathrm{ff}}^l$ coordinates, and we refer to each coordinate $z_{t,i}^l$ as an \emph{MLP channel} (the middle ``neurons'' in Figure.~\ref{fig:mlp}). Channel $i$ is formed by (a) the $i$-th row of $W_1^l$, which computes $u_{t,i}^l$, and (b) the $i$-th column of $W_2^l$, which determines how $z_{t,i}^l$ contributes to the output $y_t^l$. This channel-level view is exactly the structural unit we later keep or remove in \textsc{SecRL-Prune}.

The final output of the model depends on these MLP computations across all layers. At each time step $t$, after processing all tokens up to position $t$, the model produces a probability distribution over the vocabulary $V$ for the \emph{next} token. We denote this distribution by
\begin{equation}
p(\cdot \mid x_{<t})
= \mathrm{softmax}\!\bigl(W_{\text{out}}\, h_t^{N}\bigr),
\end{equation}
where $h_t^{N}\in\mathbb{R}^d$ is the final-layer hidden state at position $t$ and $W_{\text{out}}\in\mathbb{R}^{|V|\times d}$ is the output projection.

In practice, this distribution is typically peaked: a small set of tokens receives most of the probability mass (labeled ``Top Predictions'' in Fig.~\ref{fig:topp}), while the remaining tokens share the leftover mass (``Others'').

\subsection{KL Divergence and Knowledge Distillation}

Our pruning method (presented in Section~\ref{sec:method}) aims to compress a CodeLLM while preserving its behaviour on code-generation and mutation tasks. We therefore need a simple token-level measure of how close the pruned model’s \emph{next-token} predictions are to the original model’s predictions. We use the Kullback--Leibler (KL) divergence~\cite{kullback1951information} in a standard knowledge-distillation setup, where the original model is the teacher and the pruned model is the student.

KL divergence measures the difference between two probability distributions over the same set $V$. For distributions $p$ and $q$ over $V$, it is defined as
\begin{equation}
\label{eq:kl}
\mathrm{KL}(p \,\|\, q)
= \sum_{v \in V} p(v)\,\log\!\frac{p(v)}{q(v)} .
\end{equation}

If $\mathrm{KL}(p\,\|\,q)=0$, then $p$ and $q$ are exactly the same distribution (they assign the same probability to every token). As $\mathrm{KL}(p\,\|\,q)$ increases, $q$ deviates more from $p$, meaning the two distributions become less similar.

In our setting, at each time step $t$ (given the prefix up to $t$), the teacher defines a next-token distribution $p_T(\cdot \mid \text{prefix up to }t)$ and the student defines $p_S(\cdot \mid \text{prefix up to }t)$. The standard distillation objective averages the teacher-to-student divergence over time steps:
\begin{equation}
\label{eq:avg}
\begin{aligned}
\mathcal{L}_{\mathrm{KD}}
&= \frac{1}{T}\sum_{t=1}^{T}
\mathrm{KL}\!\Big(
p_T(\cdot \mid \text{pref } t)
\,\big\|\,
p_S(\cdot \mid \text{pref } t)
\Big).
\end{aligned}
\end{equation}

KL-based distillation is a standard technique for training compact student models to match a larger teacher~\cite{sanh2019distilbert,wang2020minilm,gu2024minillm}. In contrast,SecRL-Prune reuses this teacher-to-student KL divergence as a similarity signal that later guides our pruning policy (discussed in Section~\ref{sec:reward}).

\section{The Proposed Approach: \emph{SecRL-Prune}}
\label{sec:method}
Our overarching objective is to compress a given CodeLLM while preserving its ability to perform functionality-preserving code mutation and related adversarial code-generation tasks. To this end, we introduce \textsc{SecRL-Prune}, an RL-based structured pruning framework that operates at the level of MLP channels (defined in Section \ref{sec:mlps}) within transformer blocks. Starting from a fully trained CodeLLM (Teacher), \textsc{SecRL-Prune} learns a pruning policy that retains channels most important for code mutation while pruning others, guided by a KL-based distillation reward. We focus on pruning the feed-forward (FFN) layers because they contain the majority of model parameters (roughly 60--70\%) compared to self-attention, embeddings, and output projection layers.

\begin{figure}[t]
    \centering
    \includegraphics[
        width=\linewidth,
        trim=4cm 4.5cm 4cm 4.5cm, % left bottom right top – tweak these
        clip
    ]{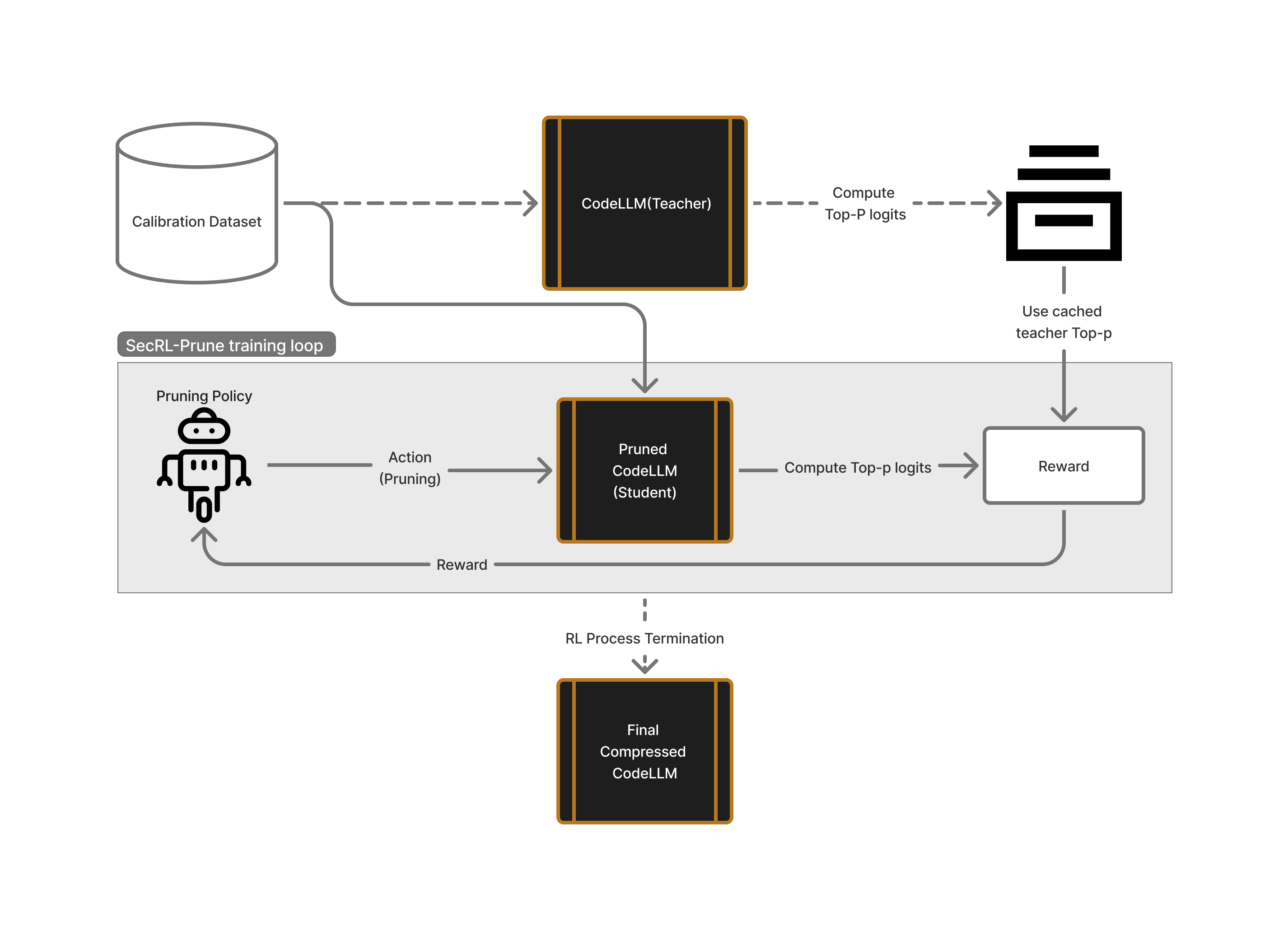}
    \caption{Overview of \textsc{SecRL-Prune} Using a calibration dataset, we run the pretrained teacher CodeLLM once to cache its top-$P$ token indices and logits. A pruning policy then proposes MLP-channel masks to form a pruned student, whose predictions on the same top-$P$ tokens are compared to the cached teacher via KL divergence to produce a reward. The reward updates the policy, and the highest-reward student is selected as the final compressed CodeLLM.}

    \label{fig:pipeline}
\end{figure}

As depicted in Figure~\ref{fig:pipeline}, \textsc{SecRL-Prune} takes two inputs: a calibration set of code prompts and a pretrained CodeLLM, which serves as the teacher. The core problem is determining, for each MLP channel (in the teacher model) across all layers, whether to keep it or remove it. This defines a combinatorially large search space (i.e., exponential in the total number of channels) that makes exhaustive or random search intractable. We formulate this as a reinforcement learning problem: rather than searching blindly, we train a policy that learns to identify which channels are important and which can be safely pruned, guided by a reward signal that measures how well each candidate mask (a binary vector keep/remove decision for every channel) preserves the teacher's behavior.

A key practical challenge is that naively comparing teacher and student outputs during training would require keeping both models in GPU memory simultaneously. We address this by running the teacher \emph{once} on the calibration set and caching only its top-$P$ token predictions at each time step, reducing memory usage significantly.

With the teacher outputs cached, we train a lightweight pruning policy that proposes, for each layer, which MLP channels to keep and which to remove (Section~\ref{sec:policy}). Each decision produces a candidate pruned model (the student), which we evaluate by comparing its predictions to the cached teacher distribution via KL divergence (Section~\ref{sec:reward}). This reward drives the policy toward masks that best preserve the teacher's output distribution. After training, we select the student achieving the highest reward as the final compressed CodeLLM (Section~\ref{sec:terms})—the policy itself is discarded, having served its purpose as a learned search mechanism.

In the following section, we give details on Top-P logit caching, the reward function, policy design, and the final output of SecRL-Prune.

\subsection{Compute Top-$P$ Logits}
\label{sec:topl}
Standard knowledge distillation requires running teacher and student models in parallel, which is memory-prohibitive for large CodeLLMs. We eliminate this bottleneck by precomputing and caching the teacher's outputs before policy training begins. Furthermore, rather than storing the full $|V|$-way distribution at each time step, we cache only the top-$P$ most likely tokens—prior work on logit-based distillation shows that this is often sufficient to match full-vocabulary distillation while being much more memory- and compute- efficient~\cite{peng2025pretraining,li2025bild,gemma2025gemma3}.

Concretely, at each time step $t$, we obtain the teacher's next-token distribution $p_T(\cdot \mid x_{<t})$, sort tokens by probability, and retain the $P$ most likely tokens, denoted $S_t$. We cache the token IDs in $S_t$ along with their teacher logits. All remaining probability mass is grouped into a single \textsc{Other} bucket:
\begin{equation}
p_T(\textsc{Other} \mid x_{<t})
= 1 - \sum_{v \in S_t} p_T(v \mid x_{<t}).
\end{equation}

\begin{figure}[t]
    \centering
    \includegraphics[
        width=0.45\linewidth,
        height=0.35\textheight,
        keepaspectratio
    ]{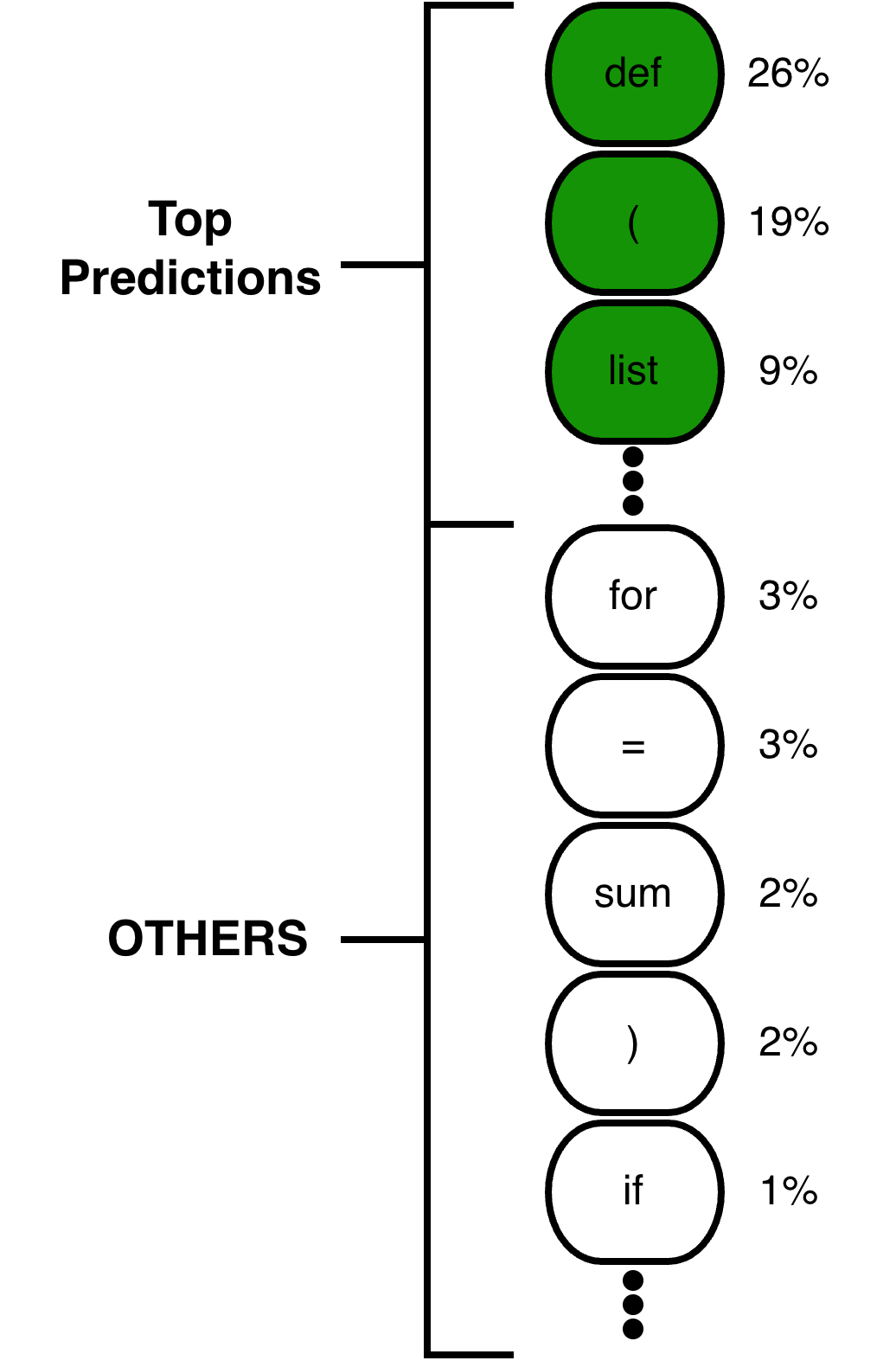}
    \caption{Output layer probabilities of the CodeLLM, grouped into Top-P tokens and an OTHER bucket.}
    \label{fig:topp}
\end{figure}

Each time step is thus represented by a compact $(P+1)$-way target: the top-$P$
tokens plus \textsc{Other} (refer to Figure~\ref{fig:topp}). For the student, we evaluate $p_S(\cdot \mid x_{<t}, m)$ on the teacher-cached top-$P$ token set $S_t$ and define the aggregated remainder mass as
$p_S(\textsc{Other}\mid x_{<t}, m) \;=\; 1 - \sum_{v \in S_t} p_S(v \mid x_{<t}, m)$.

\subsection{Policy Architecture}
\label{sec:policy}

The goal of the policy is to learn a probability distribution over pruning masks---that is, which MLP channels to keep and which to remove across all layers. Rather than enumerating pruning masks directly, we allow the system to learn per-channel scores that define this distribution: higher scores make a channel more likely to be retained. During training, the system samples masks from this distribution, evaluates the resulting student via the KL-based reward, and uses the reward signal to update the scores toward masks that better preserve the teacher's behaviour.

A standard RL approach would train a neural network that observes some representation of the model and outputs pruning decisions. We opt for a simpler design: we directly learn one parameter vector per transformer block, where each entry scores the importance of a corresponding MLP channel. Recall from Section~\ref{sec:mlps} that in layer $l$ the MLP computes an intermediate activation $z_t^l\in\mathbb{R}^{d_{\mathrm{ff}}^l}$, where each coordinate corresponds to an MLP channel.

For each layer $l$, we maintain a learnable vector $\pi^l\in\mathbb{R}^{d_{\mathrm{ff}}^l}$ with one score per channel. At each policy step, we convert these scores into a probability distribution (via softmax) and sample a fixed number of channels to \emph{keep}, matching the target pruning ratio. The selected channels form a binary mask $m^l\in\{0,1\}^{d_{\mathrm{ff}}^l}$.

During a forward pass, we apply the mask \emph{after} the activation and \emph{before} the second linear projection (see Figure~\ref{fig:mlp}). That is, after computing $z_t^l=\sigma(u_t^l)$, we gate channels by
\begin{equation}
\begin{aligned}
z_t^l &\leftarrow m^l \odot z_t^l,\\
y_t^l &= W_2^l\, z_t^l,
\end{aligned}
\end{equation}
where $\odot$ denotes element-wise multiplication.

If $m_i^l = 0$, channel $i$ contributes nothing to the MLP output (its activation is suppressed), so the corresponding structural components can be removed: the $i$-th row of $W_1^l$ and the $i$-th column of $W_2^l$. If $m_i^l = 1$, the channel is preserved. Collecting $\{\pi^l\}_{l=1}^{N}$ defines the full pruning policy across the model.

\subsection{Reward Function}
\label{sec:reward}

The reward should encourage pruning masks that preserve the teacher's output distribution. Given a mask $m$, let $p_T(\cdot \mid x_{<t})$ and $p_S(\cdot \mid x_{<t}, m)$ denote the teacher and student next-token distributions at time step $t$. We measure their divergence using KL, but rather than summing over the full vocabulary $V$ (Eq.~\ref{eq:kl}), we sum over the cached Top-$P$ set $S_t$ and add a term for the \textsc{Other} bucket (Section~\ref{sec:topl}). This yields a per-step distillation loss over a compact $(P{+}1)$ distribution:
\begin{align}
\ell_t(m)
&=
\sum_{v\in S_t} p_T(v\mid x_{<t})\,
\log\frac{p_T(v\mid x_{<t})}{p_S(v\mid x_{<t},m)}
\nonumber\\
&\quad+\;
p_T(\textsc{Other}\mid x_{<t})\,
\log\frac{p_T(\textsc{Other}\mid x_{<t})}{p_S(\textsc{Other}\mid x_{<t},m)}.
\label{eq:lt}
\end{align}

We aggregate this loss across all time steps (Eq.~\ref{eq:avg}) to obtain $\mathcal{L}_{\mathrm{KD}}(m)$ for a single code prompt, then define the reward as the negative expected distillation loss over the calibration set:
\begin{equation}
\label{eq:reward_expectation}
R(m)
= -\mathbb{E}_{\text{prompt} \in \mathcal{D}_{\text{cal}}}\!\left[\mathcal{L}_{\mathrm{KD}}(m)\right].
\end{equation}

Higher reward corresponds to masks that better preserve the teacher's distribution. We update the policy parameters (i.e., per-channel scores learned by the RL agent, discussed in Section~\ref{sec:policy}) using REINFORCE~\cite{williams1992reinforce}, a policy-gradient algorithm that adjusts the per-channel scores based on the reward signal from sampled masks.

\subsection{The output of \textsc{SecRL-Prune}}
\label{sec:terms}
After the policy-training loop terminates (refer to Figure~\ref{fig:pipeline}), we select the pruning mask $m^\star$ that achieved the highest reward on the calibration set. We then apply this mask to the teacher model, converting it into a \emph{structurally} pruned CodeLLM by retaining only the MLP channels marked active by $m^\star$ and physically removing the corresponding rows of $W_1^l$ and columns of $W_2^l$ in each layer (as discussed in Section~\ref{sec:policy}). The resulting student is a smaller, faster CodeLLM that requires no teacher cache at inference time. This pruned model (not the policy learned by the RL agent) is the sole output of \textsc{SecRL-Prune} and serves as the final compressed CodeLLM.

\section{Experimental Setup}
\label{sec:exp}
\subsection{Models and Baselines}
\label{sec:models}

We evaluate \textsc{SecRL-Prune} on three widely used CodeLLMs: CodeLlama-7B-Instruct, CodeLlama-7B-Python~\cite{roziere2023codellama}, and Qwen2.5-Coder-7B-Instruct~\cite{hui2024qwen25coder}. For all models, we use the official released checkpoints without additional finetuning.

\textsc{SecRL-Prune} operates exclusively on the FFN/MLP sub-layers, which contain the majority of model parameters. For compression ratio $\rho \in \{0.10, 0.20, 0.25, 0.30\}$, we prune a $\rho$ fraction of MLP channels in each layer while keeping attention, embeddings, and the output projection unchanged. The resulting student retains the teacher's depth and attention configuration but has reduced FFN widths, keeping a $(1-\rho)$ fraction of FFN channels per layer.

We compare against two reference points: (i) the uncompressed teacher as an upper bound, and (ii) PruneNet~\cite{sengupta2025you}, a recent state-of-the-art structured pruning method based on policy learning. We select PruneNet as our primary baseline because it has been shown to outperform earlier structured pruning methods, making it a competitive reference for evaluating structured pruning on CodeLLMs ~\cite{sengupta2025you}.

\begin{figure*}[h]
    \centering
    \begin{subfigure}[t]{\textwidth}
        \centering
        \includegraphics[width=0.85\linewidth,height=4.3cm]{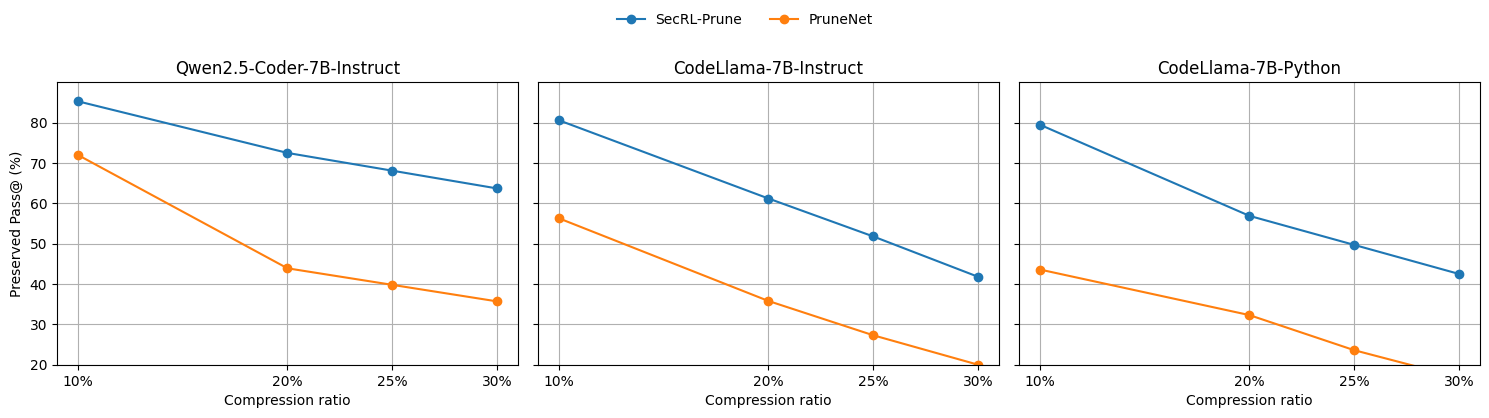}
        \caption{Preserved HumanEval correctness vs.\ compression ratio for three CodeLLMs under \textsc{SecRL-Prune} and PruneNet.}
        \label{fig:linechart-pass}
    \end{subfigure}
    \vspace{0.6em}
    \begin{subfigure}[t]{\textwidth}
        \centering
        \includegraphics[width=0.85\linewidth,height=4.3cm]{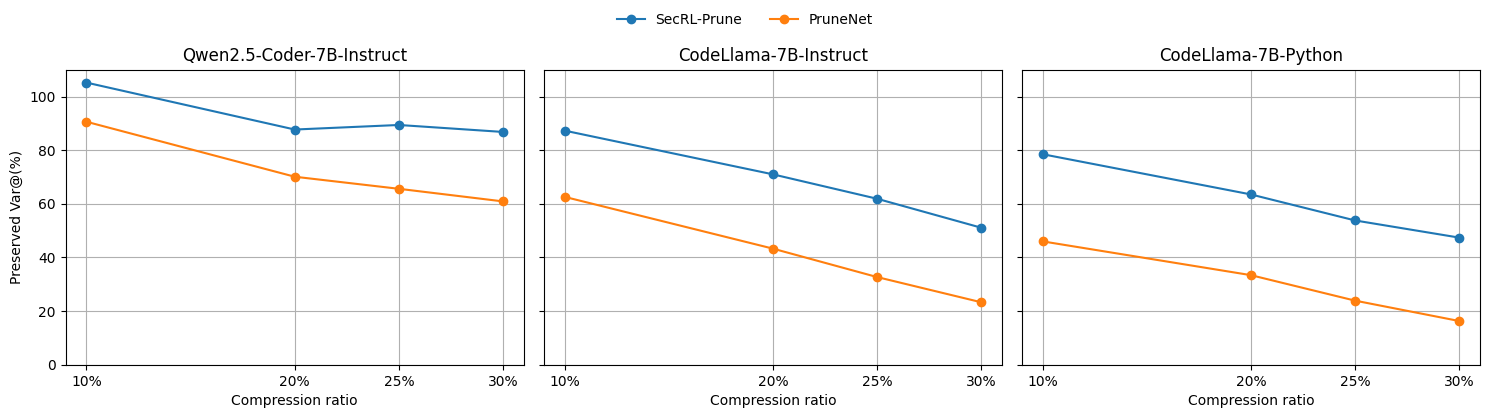}
        \caption{Preserved HumanEval variability vs.\ compression ratio for three CodeLLMs under \textsc{SecRL-Prune} and PruneNet.}
        \label{fig:linechart-var}
    \end{subfigure}

    \caption{Preservation trends on HumanEval under \textsc{SecRL-Prune} vs.\ PruneNet.}
    \label{fig:linecharts}
\end{figure*}

\subsection{Evaluation Protocol}
\label{sec:eval}

Our primary objective is to evaluate how well \textsc{SecRL-Prune} preserves the code-generation and mutation capabilities of a teacher CodeLLM after structured pruning. We use the HumanEval benchmark~\cite{chen2021codex}, which contains Python programming problems with unit tests, and evaluate pruned student models against their corresponding teachers.

We report three families of metrics. First, \textbf{pass@k}~\cite{chen2021codex} measures execution accuracy using the standard protocol and estimator for $k \in \{1, 10, 100\}$. Second, \textbf{var@k}~\cite{madani2024metamorphic} measures diversity among correct outputs on problems where the model produces at least one correct solution. For each HumanEval task, we consider the first $k$ generations, keep only those that pass the unit tests, and then count the number of unique correct solutions using hashes of the normalized code. While pass@k captures whether the model finds any correct solution, var@k captures how diverse its correct solutions are; the two should be interpreted together. Third, \textbf{Preserved correctness (Pres.)} reports the percentage of the teacher's pass@k retained by the pruned student, averaged over $k \in \{1, 10, 100\}$.

The calibration set used during pruning  contains a total of 300 prompts, sampled from MBPP ~\cite{austin2021program} and a small collection of GitHub functions. During pruning, both teacher and student are prompted only with the natural-language description and function signature; unit tests are never used in the RL loop.

\subsection{Implementation Details}
\label{sec:impl}

All experiments are implemented in PyTorch using HuggingFace Transformers. \textsc{SecRL-Prune} is trained with REINFORCE using a Top-$P$ cache size of $P = 128$, batch size of 300, learning rate of $3 \times 10^{-4}$, and 10 policy updates per compression ratio. Experiments run on NVIDIA A100 40\,GB GPUs, and we report the mean over five random seeds for each configuration.

To quantify training cost, we measure peak GPU memory usage during RL optimization (Figure~\ref{fig:barchart}). Because \textsc{SecRL-Prune} caches the teacher once and trains with only the student in memory, it uses substantially less peak memory than approaches that require both models to remain loaded throughout training.

\subsection{Real-World Malware Mutation Against VirusTotal Detectors}
To evaluate whether semantics-preserving code mutation affects malware detectability, we conducted a small case study using three real-world samples from MalwareBazaar~\cite{malwarebazaar} (two Python and one JavaScript). For each sample, we first uploaded the original file to VirusTotal~\cite{virustotal} and recorded the point-in-time detection score as ``flagging engines / total engines.'' Next, we randomly selected a single function in each sample and generated a syntactically different but semantically equivalent rewrite using a pruned CodeLLM at 20\% compression (CodeLlama-7B-Instruct or Qwen2.5-Coder-7B-Instruct), leaving the rest of the file unchanged. We then uploaded the mutated variant to VirusTotal and recorded the updated detection score under the same reporting format. We report before/after detections for each sample and mutator; results are point-in-time and intended as an illustrative case study rather than a large-scale statistical evaluation.

\begin{table*}[h]
\centering
\setlength{\tabcolsep}{3pt}
\renewcommand{\arraystretch}{0.95}
\caption{HumanEval functional correctness (pass@k) for three CodeLLMs under different structured pruning methods. For each model and compression level, “Pres.” is the average, over pass@1/10/100, of how large the pruned model’s pass@k remains relative to the dense model’s pass@k, reported as a percentage.}
\resizebox{\textwidth}{!}{%
\begin{tabular}{c l cccc cccc cccc}
\toprule
Comp. & Method &
\multicolumn{4}{c}{Qwen2.5-Coder-7B-Instruct} &
\multicolumn{4}{c}{CodeLlama-7B-Instruct} &
\multicolumn{4}{c}{CodeLlama-7B-Python} \\
 & &
\textbf{p@1} & \textbf{p@10} & \textbf{p@100} & \textbf{Pres.} &
\textbf{p@1} & \textbf{p@10} & \textbf{p@100} & \textbf{Pres.} &
\textbf{p@1} & \textbf{p@10} & \textbf{p@100} & \textbf{Pres.} \\
\midrule
0\%  & Dense &
71.57\% & 85.58\% & 89.63\% & 100.0\% &
27.67\% & 60.04\% & 83.53\% & 100.0\% &
36.70\% & 68.30\% & 87.80\% & 100.0\% \\[0.5ex]
\cdashline{1-14}\noalign{\vskip 0.5ex}
10\% & SecRL-Prune &
54.79\% & 73.18\% & 84.21\% & \textbf{85.3\%} &
21.37\% & 46.91\% & 72.25\% & \textbf{80.6\%} &
26.20\% & 53.50\% & 78.04\% & \textbf{79.5\%} \\
     & PruneNet &
46.43\% & 62.08\% & 70.34\% & 72.0\% &
14.96\% & 33.07\% & 49.85\% & 56.3\% &
14.16\% & 30.09\% & 42.40\% & 43.6\% \\[0.5ex]
\cdashline{1-14}\noalign{\vskip 0.5ex}
20\% & SecRL-Prune &
41.95\% & 60.78\% & 78.65\% & \textbf{72.5\%} &
15.06\% & 33.78\% & 60.97\% & \textbf{61.2\%} &
18.94\% & 35.93\% & 58.39\% & \textbf{56.9\%} \\
     & PruneNet &
21.30\% & 38.57\% & 51.05\% & 43.9\% &
7.98\%  & 20.50\% & 37.19\% & 35.8\% &
10.55\% & 20.77\% & 33.10\% & 32.3\% \\[0.5ex]
\cdashline{1-14}\noalign{\vskip 0.5ex}
25\% & SecRL-Prune &
38.16\% & 57.38\% & 75.14\% & \textbf{68.1\%} &
11.91\% & 27.72\% & 55.33\% & \textbf{51.8\%} &
15.32\% & 31.58\% & 53.68\% & \textbf{49.7\%} \\
     & PruneNet &
18.70\% & 34.71\% & 47.29\% & 39.8\% &
5.96\%  & 14.73\% & 29.88\% & 27.3\% &
7.90\%  & 15.04\% & 23.80\% & 23.6\% \\[0.5ex]
\cdashline{1-14}\noalign{\vskip 0.5ex}
30\% & SecRL-Prune &
34.37\% & 53.98\% & 71.63\% & \textbf{63.7\%} &
8.75\%  & 20.65\% & 49.69\% & \textbf{41.8\%} &
11.70\% & 27.23\% & 48.97\% & \textbf{42.5\%}  \\
     & PruneNet &
16.10\% & 30.85\% & 43.53\% & 35.7\% &
4.20\%  & 9.70\%  & 23.85\% & 20.0\% &
5.80\%  & 10.00\% & 17.60\% & 16.8\% \\[0.4ex]
\bottomrule
\end{tabular}}
\label{tab:humaneval_passk}
\end{table*}

\begin{table*}[t]
\centering
\setlength{\tabcolsep}{3pt}
\renewcommand{\arraystretch}{0.95}
\caption{HumanEval mutation diversity (var@k) for three CodeLLMs under different structured pruning methods. “Pres.” is the average, over var@10 and var@100, of how large the pruned model’s var@k remains relative to the dense model’s var@k, reported as a percentage.}
\begin{tabular}{c l ccc ccc ccc}
\toprule
Comp. & Method &
\multicolumn{3}{c}{Qwen2.5-Coder-7B-Instruct} &
\multicolumn{3}{c}{CodeLlama-7B-Instruct} &
\multicolumn{3}{c}{CodeLlama-7B-Python} \\
 & &
\textbf{var@10} & \textbf{var@100} & \textbf{Pres.} &
\textbf{var@10} & \textbf{var@100} & \textbf{Pres.} &
\textbf{var@10} & \textbf{var@100} & \textbf{Pres.} \\
\midrule
0\%  & Dense &
34.71\% & 12.50\% & 100.0\% &
33.89\% & 15.24\% & 100.0\% &
35.88\% & 20.06\% & 100.0\% \\[0.5ex]
\cdashline{1-11}\noalign{\vskip 0.5ex}
10\% & SecRL-Prune &
35.72\% & 13.44\% & \textbf{105.2\%} &
31.97\% & 12.22\% & \textbf{87.3\%} &
29.44\% & 15.04\% & \textbf{78.5\%} \\
     & PruneNet &
29.33\% & 12.08\% & 90.6\% &
26.62\% & 7.11\% & 62.6\% &
22.18\% & 6.05\% & 46.0\% \\[0.5ex]
\cdashline{1-11}\noalign{\vskip 0.5ex}
20\% & SecRL-Prune &
27.12\% & 12.17\% & \textbf{87.7\%} &
26.98\% & 9.50\% & \textbf{71.0\%} &
26.96\% & 10.39\% & \textbf{63.5\%} \\
     & PruneNet &
20.24\% & 10.23\% & 70.1\% &
18.80\% & 4.75\% & 43.3\% &
16.46\% & 4.19\% & 33.4\% \\[0.5ex]
\cdashline{1-11}\noalign{\vskip 0.5ex}
25\% & SecRL-Prune &
27.71\% & 12.37\% & \textbf{89.4\%} &
23.64\% & 8.25\% & \textbf{61.9\%} &
23.00\% & 8.72\% & \textbf{53.8\%} \\
     & PruneNet &
18.51\% & 9.74\% & 65.6\% &
14.27\% & 3.56\% & 32.7\% &
12.43\% & 2.63\% & 23.9\% \\[0.5ex]
\cdashline{1-11}\noalign{\vskip 0.5ex}
30\% & SecRL-Prune &
26.48\% & 12.17\% & \textbf{86.8\%} &
18.91\% & 7.08\% & \textbf{51.1\%} &
20.53\% & 7.55\% & \textbf{47.4\%} \\
     & PruneNet &
16.72\% & 9.21\% & 60.9\% &
9.83\% & 2.67\% & 23.3\% &
8.57\% & 1.75\% & 16.3\% \\[0.4ex]
\bottomrule
\end{tabular}
\label{tab:var_results}
\end{table*}

\section{Results and Discussion}
\label{sec:results}

We evaluate \textsc{SecRL-Prune} across three CodeLLMs and four compression ratios, comparing against PruneNet and the uncompressed teacher. Our results demonstrate that \textsc{SecRL-Prune} consistently preserves more of the teacher's coding correctness and code mutation than the baseline, while requiring substantially less memory during training. We discuss each finding in turn, then address limitations and directions for future work.

\subsection{Preserved Coding Capability}
\label{sec:res_accuracy}

Table~\ref{tab:humaneval_passk} summarizes HumanEval pass@k scores across all configurations. As expected, the uncompressed teacher attains the strongest performance, and all pruning methods degrade as compression increases. However, \textsc{SecRL-Prune} degrades more gracefully than the current state-of-the-art PruneNet~\cite{sengupta2025you} across the board.

At 10\% compression, \textsc{SecRL-Prune} retains roughly 80--85\% of the teacher's average pass@k, compared to 45--72\% for PruneNet. The gap widens as pruning becomes more aggressive: at 30\% compression, \textsc{SecRL-Prune} roughly doubles PruneNet's preserved correctness on Qwen2.5-Coder-7B-Instruct (63.7\% vs.\ 35.7\%) and maintains clear advantages on both CodeLlama variants. Figure~\ref{fig:linecharts} visualizes this trend: \textsc{SecRL-Prune} follows a flatter, higher preservation curve at every compression level, with the advantage most pronounced at aggressive pruning ratios.

These results indicate that KL-guided channel selection effectively identifies which FFN capacity is essential for code generation. Even at 30\% compression—where absolute pass@1 drops substantially—the pruned models still solve a non-trivial fraction of HumanEval tasks, suggesting that CodeLLMs retain considerable capability even under aggressive structured pruning.

\subsection{Preserved Code Mutation Capability}
\label{sec:res_diversity}

Beyond correctness, a mutation-capable CodeLLM must generate diverse correct variants. Table~\ref{tab:var_results} reports $\mathrm{var}@k$, which measures diversity (i.e., producing correct semantics with different syntactical structures)  among correct outputs on tasks where the model produces at least one solution.

Across all three CodeLLMs, \textsc{SecRL-Prune} preserves $\mathrm{var}@k$ substantially better than PruneNet, especially at higher compression ratios. On Qwen2.5-Coder-7B-Instruct, light pruning even slightly improves solution diversity (var@10 increases from 34.71\% to 35.72\% at 10\% compression), and var@100 remains close to the dense baseline across compression levels. In contrast, PruneNet degrades more sharply: at 30\% compression, \textsc{SecRL-Prune} retains var@100 of 7--8\% on the CodeLlama variants, whereas PruneNet collapses below 3\%.

These findings suggest that our channel selection retains not just the capacity to produce correct code, but also the variability needed for effective code mutation. However, $\mathrm{var}@k$ must be interpreted alongside pass@k: since $\mathrm{var}@k$ is computed only over solved problems, a model with low pass@k can still exhibit relatively high $\mathrm{var}@k$ on the few tasks it solves. The two metrics together provide a more complete picture of both correctness and diversity.

\subsection{Resource Efficiency and Security Implications}
\label{sec:res_resources}

\begin{figure}[t]
    \centering
    \includegraphics[width=\linewidth]{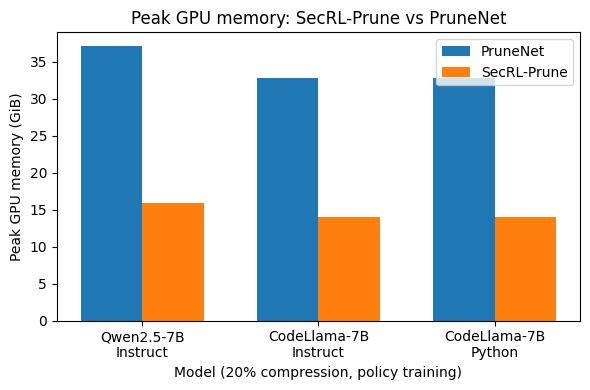}
    \caption{Peak GPU memory usage during pruning-policy training at 20\% compression: \textsc{SecRL-Prune} consistently uses less than half the memory of PruneNet.}
    \label{fig:barchart}
\end{figure}

Figure~\ref{fig:barchart} compares peak GPU memory during policy training at 20\% compression. Across all three teachers, \textsc{SecRL-Prune} uses less than half the memory of PruneNet: peak usage drops from 33--37\,GiB to 14--16\,GiB, a consistent 55--60\% reduction. This efficiency stems directly from our caching design: \textsc{SecRL-Prune} runs policy optimization with only the student in memory, whereas PruneNet requires both teacher and student to remain loaded throughout training.

From a security perspective, this lower resource footprint carries important implications. If a mutation-capable CodeLLM can be trained on a single commodity GPU and compressed to run on endpoint-class devices—exactly where many intrusions originate—it becomes more realistic to embed such models into malware or operate them on cheap, disposable infrastructure. Our results suggest that RL-based structured pruning not only preserves mutation capability, but may also lower the barrier to obtaining and deploying specialized code-mutation engines. This dual-use concern motivates continued research into detection and defense mechanisms alongside compression techniques.

\subsection{Real-World Impact: Malware Detection Evasion}
\label{sec:virustotal}

The metrics reported thus far—pass@k and $\mathrm{var}@k$—measure correctness and diversity on programming benchmarks, but they do not directly demonstrate whether pruned CodeLLMs retain practical utility for adversarial tasks. To bridge this gap, we evaluate whether mutations generated by our compressed models can evade real-world malware detection systems.

\begin{figure}[t]
    \centering
    \begin{tikzpicture}[
        box/.style={rectangle, draw, rounded corners, minimum width=2.4cm, minimum height=0.7cm, align=center, font=\footnotesize},
        inputbox/.style={box, fill=gray!15},
        processbox/.style={box, fill=blue!10},
        outputbox/.style={box, fill=green!10},
        evalbox/.style={box, fill=orange!10},
        resultbox/.style={box, fill=red!10},
        arrow/.style={->, >=stealth, thick},
        label/.style={font=\scriptsize, align=center}
    ]
    
    % Row 0: Input at top center
    \node[inputbox] (input) at (0, 0) {Original Malware Samples\\(Python, JavaScript)};
    
    % Row 1: Two parallel nodes
    \node[evalbox] (vt1) at (-1.8, -1.5) {VirusTotal};
    \node[processbox] (mutator) at (1.8, -1.5) {Pruned CodeLLM\\(20\% compression)};
    
    % Row 2: Results left, Mutated code right
    \node[resultbox] (result1) at (-1.8, -3.0) {\# of Detection (i.e., $D_{\text{org}}$)};
    \node[outputbox] (mutated) at (1.8, -3.0) {Mutated Code};
    
    % Row 3: VirusTotal for mutated
    \node[evalbox] (vt2) at (1.8, -4.5) {VirusTotal};
    
    % Row 4: Result for mutated
    \node[resultbox] (result2) at (1.8, -6.0) {\# of Detection (i.e., $D_{\text{mut}}$)};
    
    % Row 5: Comparison at bottom center
    \node[box, fill=yellow!20] (compare) at (0, -7.5) {Detection Reduction: $\Delta D$};
    
    % Arrows - left path
    \draw[arrow] (input.south) -- ++(0,-0.4) -| (vt1.north);
    \draw[arrow] (vt1) -- (result1);
    \draw[arrow] (result1.south) |- (compare.west);
    
    % Arrows - right path
    \draw[arrow] (input.south) -- ++(0,-0.4) -| (mutator.north);
    \draw[arrow] (mutator) -- (mutated);
    \draw[arrow] (mutated) -- (vt2);
    \draw[arrow] (vt2) -- (result2);
    \draw[arrow] (result2.south) |- (compare.east);
    
    % Path labels
    \node[label] at (-3.2, -2.2) {Original\\path};
    \node[label] at (3.2, -2.2) {Mutation\\path};
    
    \end{tikzpicture}
    \caption{Experimental framework for evaluating detection evasion. Original malware is submitted directly to VirusTotal (left path) and also mutated by a pruned CodeLLM before submission (right path). The detection reduction $\Delta D$ measures evasion effectiveness.}
    \label{fig:virustotal_framework}
    
\end{figure}
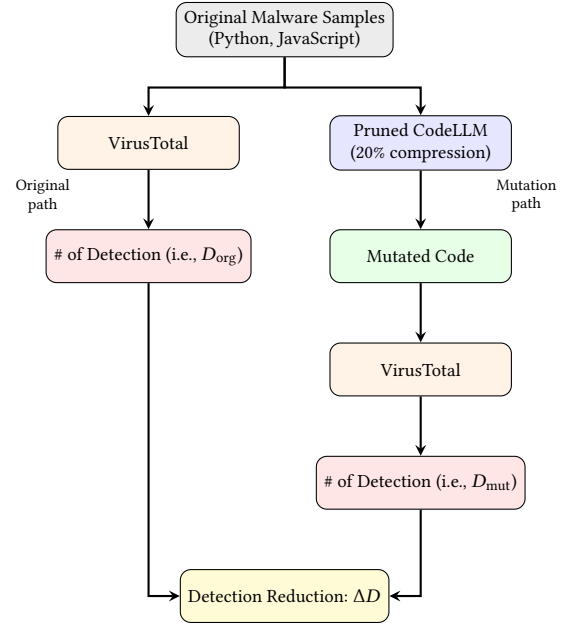

Figure~\ref{fig:virustotal_framework} illustrates our experimental setup. We selected three malware samples—two written in Python and one in JavaScript—to test generalization across languages. For each sample, we used our pruned CodeLLMs (at 20\% compression) to generate a single semantics-preserving mutation of one function, then submitted both the original and mutated versions to VirusTotal~\cite{virustotal}, which aggregates verdicts from over 60 antivirus and anti-malware engines. We compare the detection rates before and after mutation to quantify evasion effectiveness. Table~\ref{tab:virustotal} reports the results, and also includes a coarse mutation magnitude measure: \emph{Mutation Level (\%)}, defined as the line-diff ratio between the mutated file and the original file (i.e., the percentage of lines added, removed, or modified relative to the original file). In our experiments, we also verified that the mutated samples remained executable and functionally correct.

The results are striking. For the first Python sample, detections drop from 28/63 to 12/63 (mutated using pruned CodeLlama) and 9/63 (mutated using pruned Qwen)—a 57--68\% reduction in detection rate from a single function rewrite. The second Python sample shows an even larger effect, falling from 25/63 to 7/63 and 6/63 (72--76\% reduction). Most dramatic is the JavaScript sample, where detections plummet from 9/62 to just 1/62 for both mutators—an 88\% reduction. Because files are small and the mutated function can occupy a large fraction of total lines, this coarse line-diff ratio can appear high even when only one function is rewritten.

These findings carry two important implications. First, they validate that our pruned CodeLLMs retain meaningful mutation capability beyond what benchmark metrics alone can capture: a model compressed by 20\% can still generate rewrites (i.e., mutate) that substantially degrade signature- and pattern-based detection. Second, the non-zero residual detections indicate that some engines employ more robust features beyond superficial syntax, suggesting a path forward for defenders. Overall, this experiment demonstrates that the security concerns motivating our work are not hypothetical—compressed CodeLLMs pose a realistic threat to static malware defenses.

\begin{table}[h]
\centering
\footnotesize
\setlength{\tabcolsep}{10pt}         % looser columns (was 4pt)
\renewcommand{\arraystretch}{1.05}  % looser rows (was 0.88)
% keep these normal (remove extreme tightening)
%\setlength{\aboverulesep}{0pt}
%\setlength{\belowrulesep}{0pt}

\caption{VirusTotal detections (flagging engines / total) before vs.\ after semantics-preserving mutation by pruned CodeLLMs at 20\% compression (point-in-time). A higher reduction in detection is better.}
\label{tab:virustotal}

\begin{tabular}{@{}l p{1.5cm} c c c@{}}
\toprule
{\bf Sample} & {\bf Variant / Mutator} & {\bf \makecell{Mutation \\ Level \%}}& {\bf \makecell{Detection}}  & \makecell{\bf Change in \\\bf Detection} \\
\midrule

\multicolumn{4}{@{}l@{}}{\textbf{Anubis Family Sample 1(Python)} (SHA-256: \texttt{1c486702...2c98a})} \\
Original & -- & 0\% &28/63 \\
Mutated  & CodeLlama-7B-Instruct (20\% pruned)   & 8\% &12/63 & $\downarrow 57\%$\\
Mutated  & Qwen2.5-Coder-7B-Instruct (20\% pruned)  & 60\% &9/63 & $\downarrow 68\%$\\
\specialrule{0.8pt}{1.2pt}{1.2pt}

\multicolumn{4}{@{}l@{}}{\textbf{Anubis Family Sample 2(Python)} (SHA-256: \texttt{500c9ec9...2cb42})} \\
Original & -- & 0\% & 25/63 \\
Mutated  & CodeLlama-7B-Instruct (20\% pruned)  & 18\% &7/63 & $\downarrow 72\%$ \\
Mutated  & Qwen2.5-Coder-7B-Instruct (20\% pruned) & 50\% &6/63 & $\downarrow 76\%$ \\
\specialrule{0.8pt}{1.2pt}{1.2pt}

\multicolumn{4}{@{}l@{}}{\textbf{SilentStealer Sample (Javascript)} (SHA-256: \texttt{79f8b6b8...3b55d})} \\
Original & --  & 0\% & 9/62 \\
Mutated  & CodeLlama-7B-Instruct (20\% pruned)  & 5\% &1/62  & $\downarrow 88\%$\\
Mutated  & Qwen2.5-Coder-7B-Instruct (20\% pruned)  & 11\% &1/62  & $\downarrow 88\%$\\
\bottomrule
\end{tabular}
\end{table}

\section{Conclusion and Future Work}
\label{sec:conclusion}

This paper studied an emerging security risk: whether CodeLLMs can be structurally compressed while retaining the functionality-preserving code mutation behavior that enables rapid generation of malware variants and evasion of static defenses. We introduced \textsc{SecRL-Prune}, an RL-based structured pruning framework that treats channel selection as a learned search problem using a KL-guided teacher--student reward, together with a Top-$P$ caching mechanism that reduces peak GPU memory by over 50\% compared to prior methods.

Across three real-world CodeLLMs and four compression ratios, \textsc{SecRL-Prune} consistently preserved more execution-level correctness (pass@k) and mutation diversity ($\mathrm{var}@k$) than a strong policy-learning baseline. These results show that code-mutation capability can survive significant structured compression, increasing the feasibility of compact mutation-focused models while helping defenders assess the risk posed by miniaturized mutation engines.

Several limitations point toward future work: the reward is defined over a calibration distribution and truncated Top-$P$ support, which may under-emphasize rare or long-tail behaviors; $\mathrm{var}@k$ captures diversity only among correct outputs; and we prune only FFN channels while leaving attention untouched. Future work should explore broader calibration data, alternative diversity metrics, joint pruning of FFN and attention components, and staged prune-and-recover pipelines with intermediate fine-tuning.

To our knowledge, this is the first structured pruning framework tailored specifically to CodeLLMs. While the pruned models remain meaningfully capable, a noticeable gap to the dense teacher persists, motivating broader validation across additional models, benchmarks, and programming languages.

\bibliographystyle{ACM-Reference-Format}
\balance
\bibliography{sample-base}


\begin{thebibliography}{29}

%%% the Web address.

\ifx \showCODEN    \undefined \def \showCODEN     #1{\unskip}     \fi
\ifx \showISBNx    \undefined \def \showISBNx     #1{\unskip}     \fi
\ifx \showISBNxiii \undefined \def \showISBNxiii  #1{\unskip}     \fi
\ifx \showISSN     \undefined \def \showISSN      #1{\unskip}     \fi
\ifx \showLCCN     \undefined \def \showLCCN      #1{\unskip}     \fi
\ifx \shownote     \undefined \def \shownote      #1{#1}          \fi
\ifx \showarticletitle \undefined \def \showarticletitle #1{#1}   \fi
\ifx \showURL      \undefined \def \showURL       {\relax}        \fi
% invisible to TeX
\providecommand\bibfield[2]{#2}
\providecommand\bibinfo[2]{#2}
\providecommand\natexlab[1]{#1}
\providecommand\showeprint[2][]{arXiv:#2}

\bibitem[{abuse.ch}({[n.\,d.]})]%
        {malwarebazaar}
\bibfield{author}{\bibinfo{person}{{abuse.ch}}.}
  \bibinfo{year}{[n.\,d.]}\natexlab{}.
\newblock \bibinfo{title}{{MalwareBazaar}}.
\newblock \bibinfo{howpublished}{Website}.
\newblock
\newblock
\shownote{\url{https://bazaar.abuse.ch/}}.


\bibitem[Anthropic(2025)]%
        {anthropic2025aiespionage}
\bibfield{author}{\bibinfo{person}{Anthropic}.}
  \bibinfo{year}{2025}\natexlab{}.
\newblock \showarticletitle{Disrupting the first reported {AI}-orchestrated
  cyber espionage campaign}.
\newblock \bibinfo{journal}{\emph{Anthropic News}} (\bibinfo{date}{13 November}
  \bibinfo{year}{2025}).
\newblock
\urldef\tempurl%
\url{https://www.anthropic.com/news/disrupting-AI-espionage}
\showURL{%
\tempurl}
\newblock
\shownote{Accessed: 2025-12-10}.


\bibitem[Austin et~al\mbox{.}(2021)]%
        {austin2021program}
\bibfield{author}{\bibinfo{person}{Jacob Austin}, \bibinfo{person}{Augustus
  Odena}, \bibinfo{person}{Maxwell Nye}, \bibinfo{person}{Maarten Bosma},
  {et~al\mbox{.}}} \bibinfo{year}{2021}\natexlab{}.
\newblock \showarticletitle{Program Synthesis with Large Language Models}.
\newblock \bibinfo{journal}{\emph{arXiv preprint arXiv:2108.07732}}
  (\bibinfo{year}{2021}).
\newblock
\urldef\tempurl%
\url{https://arxiv.org/abs/2108.07732}
\showURL{%
\tempurl}


\bibitem[Bensaoud et~al\mbox{.}(2024)]%
        {bensaoud2024survey}
\bibfield{author}{\bibinfo{person}{Ahmed Bensaoud}, \bibinfo{person}{Jugal
  Kalita}, {and} \bibinfo{person}{Mahmoud Bensaoud}.}
  \bibinfo{year}{2024}\natexlab{}.
\newblock \showarticletitle{A Survey of Malware Detection Using Deep Learning
  (Static, Dynamic, Hybrid)}.
\newblock \bibinfo{journal}{\emph{Machine Learning with Applications}}
  \bibinfo{volume}{16} (\bibinfo{year}{2024}), \bibinfo{pages}{100546}.
\newblock
\href{https://doi.org/10.1016/j.mlwa.2024.100546}{doi:\nolinkurl{10.1016/j.mlwa.2024.100546}}


\bibitem[Chen et~al\mbox{.}(2021)]%
        {chen2021codex}
\bibfield{author}{\bibinfo{person}{Mark Chen}, \bibinfo{person}{Jerry Tworek},
  \bibinfo{person}{Heewoo Jun}, \bibinfo{person}{Qiming Yuan}, {and}
  \bibinfo{person}{et. al.}} \bibinfo{year}{2021}\natexlab{}.
\newblock \showarticletitle{Evaluating Large Language Models Trained on Code}.
\newblock  (\bibinfo{year}{2021}).
\newblock
\showeprint[arxiv]{2107.03374}~[cs.LG]


\bibitem[Dettmers et~al\mbox{.}(2023)]%
        {dettmers2023qlora}
\bibfield{author}{\bibinfo{person}{Tim Dettmers}, \bibinfo{person}{Artidoro
  Pagnoni}, \bibinfo{person}{Ari Holtzman}, {and} \bibinfo{person}{Luke
  Zettlemoyer}.} \bibinfo{year}{2023}\natexlab{}.
\newblock \showarticletitle{{QLoRA}: Efficient Finetuning of Quantized {LLMs}}.
  In \bibinfo{booktitle}{\emph{Advances in Neural Information Processing
  Systems (NeurIPS)}}.
\newblock
\urldef\tempurl%
\url{https://arxiv.org/abs/2305.14314}
\showURL{%
\tempurl}
\newblock
\shownote{arXiv preprint arXiv:2305.14314}.


\bibitem[Frantar and Alistarh(2023)]%
        {frantar2023sparsegpt}
\bibfield{author}{\bibinfo{person}{Elias Frantar} {and} \bibinfo{person}{Dan
  Alistarh}.} \bibinfo{year}{2023}\natexlab{}.
\newblock \showarticletitle{Sparse{GPT}: Massive Language Models Can Be
  Accurately Pruned in One-Shot}. In \bibinfo{booktitle}{\emph{Proceedings of
  the 40th International Conference on Machine Learning (ICML)}}
  \emph{(\bibinfo{series}{Proceedings of Machine Learning Research},
  Vol.~\bibinfo{volume}{202})}. \bibinfo{pages}{10323--10337}.
\newblock
\urldef\tempurl%
\url{https://arxiv.org/abs/2301.00774}
\showURL{%
\tempurl}


\bibitem[{Gemma Team}(2025)]%
        {gemma2025gemma3}
\bibfield{author}{\bibinfo{person}{{Gemma Team}}.}
  \bibinfo{year}{2025}\natexlab{}.
\newblock \showarticletitle{Gemma 3 Technical Report}.
\newblock \bibinfo{journal}{\emph{arXiv preprint arXiv:2503.19786}}
  (\bibinfo{year}{2025}).
\newblock
\urldef\tempurl%
\url{https://arxiv.org/abs/2503.19786}
\showURL{%
\tempurl}


\bibitem[Gu et~al\mbox{.}(2023)]%
        {gu2024minillm}
\bibfield{author}{\bibinfo{person}{Yuxian Gu}, \bibinfo{person}{Li Dong},
  \bibinfo{person}{Furu Wei}, {and} \bibinfo{person}{Minlie Huang}.}
  \bibinfo{year}{2023}\natexlab{}.
\newblock \showarticletitle{{MiniLLM}: Knowledge Distillation of Large Language
  Models}.
\newblock \bibinfo{journal}{\emph{arXiv preprint arXiv:2311.13874}}
  (\bibinfo{year}{2023}).
\newblock
\urldef\tempurl%
\url{https://arxiv.org/abs/2311.13874}
\showURL{%
\tempurl}
\newblock
\shownote{Appeared at ICLR 2024}.


\bibitem[Hui et~al\mbox{.}(2024)]%
        {hui2024qwen25coder}
\bibfield{author}{\bibinfo{person}{Binyuan Hui}, \bibinfo{person}{Fan Yang},
  \bibinfo{person}{Yuxuan Ye}, {et~al\mbox{.}}}
  \bibinfo{year}{2024}\natexlab{}.
\newblock \showarticletitle{Qwen2.5-Coder Technical Report}.
\newblock \bibinfo{journal}{\emph{arXiv preprint arXiv:2409.12186}}
  (\bibinfo{year}{2024}).
\newblock
\urldef\tempurl%
\url{https://arxiv.org/abs/2409.12186}
\showURL{%
\tempurl}


\bibitem[Konstantinou(2008)]%
        {konstantinou2008metamorphic}
\bibfield{author}{\bibinfo{person}{Evangelos Konstantinou}.}
  \bibinfo{year}{2008}\natexlab{}.
\newblock \bibinfo{booktitle}{\emph{Metamorphic Virus: Analysis and
  Detection}}.
\newblock \bibinfo{type}{{T}echnical {R}eport} RHUL-MA-2008-02.
  \bibinfo{institution}{Royal Holloway, University of London}.
\newblock


\bibitem[Kullback and Leibler(1951)]%
        {kullback1951information}
\bibfield{author}{\bibinfo{person}{Solomon Kullback} {and}
  \bibinfo{person}{Richard~A. Leibler}.} \bibinfo{year}{1951}\natexlab{}.
\newblock \showarticletitle{On Information and Sufficiency}.
\newblock \bibinfo{journal}{\emph{Annals of Mathematical Statistics}}
  \bibinfo{volume}{22}, \bibinfo{number}{1} (\bibinfo{year}{1951}),
  \bibinfo{pages}{79--86}.
\newblock


\bibitem[Lemos(2025)]%
        {lemos2025llmmalware}
\bibfield{author}{\bibinfo{person}{Robert Lemos}.}
  \bibinfo{year}{2025}\natexlab{}.
\newblock \showarticletitle{How Malware Authors Are Incorporating {LLMs} to
  Evade Detection}.
\newblock \bibinfo{journal}{\emph{Dark Reading}} (\bibinfo{date}{26 November}
  \bibinfo{year}{2025}).
\newblock
\urldef\tempurl%
\url{https://www.darkreading.com/threat-intelligence/malware-authors-incorporate-llms-evade-detection}
\showURL{%
\tempurl}
\newblock
\shownote{Accessed: 2025-12-10}.


\bibitem[Li et~al\mbox{.}(2025)]%
        {li2025bild}
\bibfield{author}{\bibinfo{person}{Ming Li}, \bibinfo{person}{Fan Zhou}, {and}
  \bibinfo{person}{Xia Song}.} \bibinfo{year}{2025}\natexlab{}.
\newblock \showarticletitle{BiLD: Bi-directional Logits Difference Loss for
  Large Language Model Distillation}. In \bibinfo{booktitle}{\emph{Proceedings
  of the 31st International Conference on Computational Linguistics (COLING
  2025)}}.
\newblock


\bibitem[Liu et~al\mbox{.}(2025)]%
        {liu2025rap}
\bibfield{author}{\bibinfo{person}{Huan Liu}, \bibinfo{person}{Chenyang Tian},
  \bibinfo{person}{Xiyu Wei}, \bibinfo{person}{Jiayi Dai},
  \bibinfo{person}{Qiang Liu}, \bibinfo{person}{Tao Wei}, \bibinfo{person}{Qian
  Li}, {and} \bibinfo{person}{Lin Li}.} \bibinfo{year}{2025}\natexlab{}.
\newblock \showarticletitle{{RAP}: Runtime-Adaptive Pruning for {LLM}
  Inference}.
\newblock \bibinfo{journal}{\emph{arXiv preprint arXiv:2505.17138}}
  (\bibinfo{year}{2025}).
\newblock
\urldef\tempurl%
\url{https://arxiv.org/abs/2505.17138}
\showURL{%
\tempurl}


\bibitem[Ma et~al\mbox{.}(2023)]%
        {ma2023llmpruner}
\bibfield{author}{\bibinfo{person}{Xinyin Ma}, \bibinfo{person}{Gongfan Fang},
  {and} \bibinfo{person}{Xinchao Wang}.} \bibinfo{year}{2023}\natexlab{}.
\newblock \showarticletitle{{LLM}-Pruner: On the Structural Pruning of Large
  Language Models}. In \bibinfo{booktitle}{\emph{Advances in Neural Information
  Processing Systems 36 (NeurIPS 2023)}}.
\newblock
\urldef\tempurl%
\url{https://arxiv.org/abs/2305.11627}
\showURL{%
\tempurl}


\bibitem[Madani(2024)]%
        {madani2024metamorphic}
\bibfield{author}{\bibinfo{person}{Pooria Madani}.}
  \bibinfo{year}{2024}\natexlab{}.
\newblock \showarticletitle{Metamorphic Malware Evolution: The Potential and
  Peril of Large Language Models}.
\newblock \bibinfo{journal}{\emph{CoRR}}  \bibinfo{volume}{abs/2410.23894}
  (\bibinfo{year}{2024}).
\newblock
\urldef\tempurl%
\url{https://arxiv.org/abs/2410.23894}
\showURL{%
\tempurl}
\newblock
\shownote{Also in Proc. 5th IEEE Int. Conf. on Trust, Privacy and Security in
  Intelligent Systems and Applications (TPS-ISA), 2023}.


\bibitem[Madou et~al\mbox{.}(2006)]%
        {madou2006software}
\bibfield{author}{\bibinfo{person}{Matias Madou}, \bibinfo{person}{Bertrand
  Anckaert}, \bibinfo{person}{Patrick Moseley}, \bibinfo{person}{Saumya
  Debray}, \bibinfo{person}{Bjorn De~Sutter}, {and} \bibinfo{person}{Koen
  De~Bosschere}.} \bibinfo{year}{2006}\natexlab{}.
\newblock \showarticletitle{Software Protection Through Dynamic Code Mutation}.
  In \bibinfo{booktitle}{\emph{Information Security Applications (WISA 2005)}}
  \emph{(\bibinfo{series}{Lecture Notes in Computer Science},
  Vol.~\bibinfo{volume}{3786})}. \bibinfo{publisher}{Springer},
  \bibinfo{pages}{194--206}.
\newblock
\href{https://doi.org/10.1007/11604938_16}{doi:\nolinkurl{10.1007/11604938_16}}


\bibitem[Peng et~al\mbox{.}(2025)]%
        {peng2025pretraining}
\bibfield{author}{\bibinfo{person}{Huan Peng}, \bibinfo{person}{Xiang Lv},
  \bibinfo{person}{Yuxiang Bai}, \bibinfo{person}{Zhewei Yao},
  \bibinfo{person}{Jing Zhang}, \bibinfo{person}{Lei Hou}, {and}
  \bibinfo{person}{Juanzi Li}.} \bibinfo{year}{2025}\natexlab{}.
\newblock \showarticletitle{Pre-training Distillation for Large Language
  Models: A Design Space Exploration}. In \bibinfo{booktitle}{\emph{Proceedings
  of the 63rd Annual Meeting of the Association for Computational Linguistics
  (ACL 2025, Long Papers)}}. \bibinfo{address}{Vienna, Austria},
  \bibinfo{pages}{3603--3618}.
\newblock


\bibitem[Rozi{\`e}re et~al\mbox{.}(2023)]%
        {roziere2023codellama}
\bibfield{author}{\bibinfo{person}{Baptiste Rozi{\`e}re},
  \bibinfo{person}{Jonas Gehring}, \bibinfo{person}{Fabian Gloeckle},
  \bibinfo{person}{Sten Sootla}, \bibinfo{person}{Itai Gat},
  \bibinfo{person}{Xiaoqing~Ellen Tan}, {et~al\mbox{.}}}
  \bibinfo{year}{2023}\natexlab{}.
\newblock \showarticletitle{Code Llama: Open Foundation Models for Code}.
\newblock \bibinfo{journal}{\emph{arXiv preprint arXiv:2308.12950}}
  (\bibinfo{year}{2023}).
\newblock
\urldef\tempurl%
\url{https://arxiv.org/abs/2308.12950}
\showURL{%
\tempurl}


\bibitem[Sanh et~al\mbox{.}(2019)]%
        {sanh2019distilbert}
\bibfield{author}{\bibinfo{person}{Victor Sanh}, \bibinfo{person}{Lysandre
  Debut}, \bibinfo{person}{Julien Chaumond}, {and} \bibinfo{person}{Thomas
  Wolf}.} \bibinfo{year}{2019}\natexlab{}.
\newblock \showarticletitle{Distil{BERT}, a Distilled Version of {BERT}:
  Smaller, Faster, Cheaper and Lighter}.
\newblock \bibinfo{journal}{\emph{arXiv preprint arXiv:1910.01108}}
  (\bibinfo{year}{2019}).
\newblock
\urldef\tempurl%
\url{https://arxiv.org/abs/1910.01108}
\showURL{%
\tempurl}


\bibitem[Sengupta et~al\mbox{.}(2025)]%
        {sengupta2025you}
\bibfield{author}{\bibinfo{person}{Ayan Sengupta}, \bibinfo{person}{Siddhant
  Chaudhary}, {and} \bibinfo{person}{Tanmoy Chakraborty}.}
  \bibinfo{year}{2025}\natexlab{}.
\newblock \showarticletitle{You Only Prune Once: Designing Calibration-Free
  Model Compression with Policy Learning}. In
  \bibinfo{booktitle}{\emph{Proceedings of the Thirteenth International
  Conference on Learning Representations (ICLR)}}.
\newblock
\urldef\tempurl%
\url{https://doi.org/10.48550/arXiv.2501.15296}
\showURL{%
\tempurl}


\bibitem[Setak and Madani(2024)]%
        {setak2024finetuning}
\bibfield{author}{\bibinfo{person}{Mohammad Setak} {and}
  \bibinfo{person}{Pooria Madani}.} \bibinfo{year}{2024}\natexlab{}.
\newblock \showarticletitle{Fine-tuning {LLMs} for Code Mutation: A New Era of
  Cyber Threats}. In \bibinfo{booktitle}{\emph{2024 IEEE 6th International
  Conference on Trust, Privacy and Security in Intelligent Systems and
  Applications (TPS-ISA)}}. IEEE.
\newblock


\bibitem[{VirusTotal}({[n.\,d.]})]%
        {virustotal}
\bibfield{author}{\bibinfo{person}{{VirusTotal}}.}
  \bibinfo{year}{[n.\,d.]}\natexlab{}.
\newblock \bibinfo{title}{{VirusTotal}}.
\newblock \bibinfo{howpublished}{Website}.
\newblock
\newblock
\shownote{\url{https://www.virustotal.com/}}.


\bibitem[Wang et~al\mbox{.}(2019)]%
        {wang2019metahunt}
\bibfield{author}{\bibinfo{person}{Lin Wang}, \bibinfo{person}{Danfeng Xu},
  \bibinfo{person}{Jiang Ming}, \bibinfo{person}{Yue Fu}, {and}
  \bibinfo{person}{Dinghao Wu}.} \bibinfo{year}{2019}\natexlab{}.
\newblock \showarticletitle{{MetaHunt}: Towards Taming Malware Mutation via
  Studying the Evolution of Metamorphic Virus}. In
  \bibinfo{booktitle}{\emph{Proceedings of the 3rd Software Protection Workshop
  (SPRO'19)}}.
\newblock


\bibitem[Wang et~al\mbox{.}(2020)]%
        {wang2020minilm}
\bibfield{author}{\bibinfo{person}{Wenhui Wang}, \bibinfo{person}{Furu Wei},
  \bibinfo{person}{Li Dong}, \bibinfo{person}{Hangbo Bao}, \bibinfo{person}{Nan
  Yang}, {and} \bibinfo{person}{Ming Zhou}.} \bibinfo{year}{2020}\natexlab{}.
\newblock \showarticletitle{{MiniLM}: Deep Self-Attention Distillation for
  Task-Agnostic Compression of Pre-Trained Transformers}. In
  \bibinfo{booktitle}{\emph{Advances in Neural Information Processing Systems
  33 (NeurIPS 2020)}}.
\newblock


\bibitem[Wei et~al\mbox{.}(2024)]%
        {wei2024effectively}
\bibfield{author}{\bibinfo{person}{Xiyu Wei}, \bibinfo{person}{Yiming Li},
  \bibinfo{person}{Liang Zhao}, {and} \bibinfo{person}{Xiang Ren}.}
  \bibinfo{year}{2024}\natexlab{}.
\newblock \showarticletitle{Effectively Training {LLMs} with Structured
  Feedforward Layers}. In \bibinfo{booktitle}{\emph{Advances in Neural
  Information Processing Systems}}.
\newblock


\bibitem[Williams(1992)]%
        {williams1992reinforce}
\bibfield{author}{\bibinfo{person}{Ronald~J. Williams}.}
  \bibinfo{year}{1992}\natexlab{}.
\newblock \showarticletitle{Simple Statistical Gradient-Following Algorithms
  for Connectionist Reinforcement Learning}.
\newblock \bibinfo{journal}{\emph{Machine Learning}} \bibinfo{volume}{8},
  \bibinfo{number}{3--4} (\bibinfo{date}{May} \bibinfo{year}{1992}),
  \bibinfo{pages}{229--256}.
\newblock
\href{https://doi.org/10.1007/BF00992696}{doi:\nolinkurl{10.1007/BF00992696}}


\bibitem[Wong and Stamp(2006)]%
        {wong2006hunting}
\bibfield{author}{\bibinfo{person}{Wing Wong} {and} \bibinfo{person}{Mark
  Stamp}.} \bibinfo{year}{2006}\natexlab{}.
\newblock \showarticletitle{Hunting for Metamorphic Engines}.
\newblock \bibinfo{journal}{\emph{Journal in Computer Virology}}
  \bibinfo{volume}{2}, \bibinfo{number}{3} (\bibinfo{year}{2006}),
  \bibinfo{pages}{211--229}.
\newblock
\href{https://doi.org/10.1007/s11416-006-0018-1}{doi:\nolinkurl{10.1007/s11416-006-0018-1}}


\end{thebibliography}

\end{document}